\documentclass[aps, twocolumn, amssymb, amsmath, nobibnotes, prl, superscriptaddress]{revtex4-2}
\usepackage[ascii]{inputenc}
\usepackage{amsmath}
\usepackage{amssymb}
\usepackage{graphicx}
\usepackage{bm}
\usepackage{amsfonts}
\usepackage[T1]{fontenc}
\usepackage[bookmarks=false,linkcolor=blue,urlcolor=blue,colorlinks,citecolor=blue]{hyperref}

\begin{document}

\title{Coulomb drag and heat transfer in strange metals}

\author{A. L. Chudnovskiy}
\affiliation{I. Institut f\"ur Theoretische Physik, Universit\"at Hamburg, Notkestra\ss e 9, D-22607 Hamburg, Germany}

\author{Alex Levchenko}
\affiliation{Department of Physics, University of Wisconsin-Madison, Madison, Wisconsin 53706, USA}

\author{Alex Kamenev}
\affiliation{School of Physics and Astronomy, University of Minnesota, Minneapolis, Minnesota 55455, USA}
\affiliation{William I. Fine Theoretical Physics Institute, University of Minnesota, Minneapolis, Minnesota 55455, USA}

\date{\today}

\begin{abstract} 
We address  Coulomb drag  and near-field heat transfer in a double-layer system of incoherent metals. Each layer is modeled by an array of tunnel-coupled SYK dots with random inter-layer interactions. Depending on the strength of intra-dot interactions and inter-dot tunneling, this model captures the crossover from the Fermi liquid to a strange metal phase.  The absence of quasiparticles in the strange metal leads to temperature-independent drag resistivity, which is in strong contrast with the quadratic temperature dependence in the Fermi liquid regime. We show that all the parameters can be independently measured in near-field heat transfer experiments, performed in Fermi liquid and strange metal regimes.
\end{abstract}
\maketitle

The electronic double layers -- spatially separated and interactively coupled conducting circuits -- provide a versatile array of low-dimensional quantum systems designed to directly probe electronic correlations via non-local transport measurements such as Coulomb drag \cite{CD-Review}. Such double layers can be formed out of 0D quantum dots and point contacts \cite{Onac2006,Khrapai2006,Khrapai2007}, 1D nanowires \cite{Yamamoto2006,Laroche2011,Laroche2014,Das2020} and topological edge states \cite{Du2021}, and bilayers of 2DEG \cite{Gramila1991,Pillarisetty2002,Kellogg2003} or graphene \cite{Gorbachev2012,Dean2016,Kim2017}. These devices enable the exploration of various electron transport regimes and the identification of correlated electronic phases from the distinct temperature dependence of the drag resistance. 

In the Fermi liquid (FL) regime the drag resistance is expected to scale quadratically with the temperature at the lowest temperatures. This result follows from the simple argument of the phase space available for the quasiparticle scattering that can be accurately established in the microscopic kinetic theory \cite{Laikhtman1990,Maslov1992,Smith1993}. The interplay of screening and diffusion leads to the enhancement of drag resistance in the disordered systems \cite{Zheng1993,Kamenev1995,Flensberg1995}. At intermediate temperatures, dragging is dominated by the collective modes and resistance peaks at the energies of plasmons in 2D bilayers. The further fall-off of drag resistance at higher temperatures can be described by hydrodynamic effects and is governed by the electron liquid viscosity in clean systems \cite{Apostolov2014,Chen2015}. All these features are well understood and rigorously described within the framework of the Fermi liquid theory.    

There are known examples of essentially non-Fermi liquid behavior in systems where the quasiparticle concept breaks down. For instance, in Luttinger liquids kinematics of 1D collisions of electrons with linear spectrum dictates that drag is dominated by the inter-wire backscattering \cite{Klesse2000,Fiete2006}. This translates into the signature power-law temperature dependence of drag resistance with the power exponent dependent on the strength of electron interaction. At the lowest temperatures, however, the trans-resistivity diverges, due to the formation of locked charge density waves. The enhancement of resistance occurs also in 2D layers provided that interactions are sufficiently strong and the electron system is on one of the possible microemulsion phases at the onset of Wigner crystallization \cite{Spivak2005}. Another notable example is the regime of drag between fractional quantum Hall liquids, where the trans-resistance is determined by the scattering and Coulomb screening effects of composite fermions \cite{Ussishkin1997,KIm1997,Kim1999,Patel2017}. Ultimately, the strong coupling limit may lead to pairing and inter-layer (indirect excitonic) superfluidity \cite{Eisenstein2004,Eisenstein2014} that can be detected in the Coulomb drag counterflow setup. 

In recent years much of the attention in the context of electronic transport is devoted to understanding the strange metal (SM) behavior in strongly correlated materials with and without quasiparticles, revealing the microscopic origin of the Planckian dissipation \cite{Hartnoll:2015vb,davison2017thermoelectric,Patel2019,Hartnoll2022,PhysRevLett.124.076801,hidden-met-ins2021}. This broad interest facilitates the development of the corresponding transport theory for strange metal bilayers that may provide additional insights into the intricate physical properties of these systems. For that purpose, we use the paradigmatic Sachdev-Ye-Kitaev (SYK) model \cite{Sachdev1993,Kitaev2015,Chowdhury2022}, which describes a strongly interacting quantum many-body system without quasiparticle excitations that is maximally chaotic, nearly conformally invariant, and exactly solvable in the limit of a large number of interacting particles. We derive analytical results for the drag resistance and near-field thermal conductance in bilayers of SYK arrays. Our analysis leads us to drastically different conclusions concerning the temperature dependence of the drag resistance in the SM phase as compared to the FL, and different from the earlier study based on the hydrodynamic-like holographic model of the strange metal \cite{Stoof}.  

 \begin{figure}[t!]
  \centering
  \includegraphics[width=0.5\textwidth]{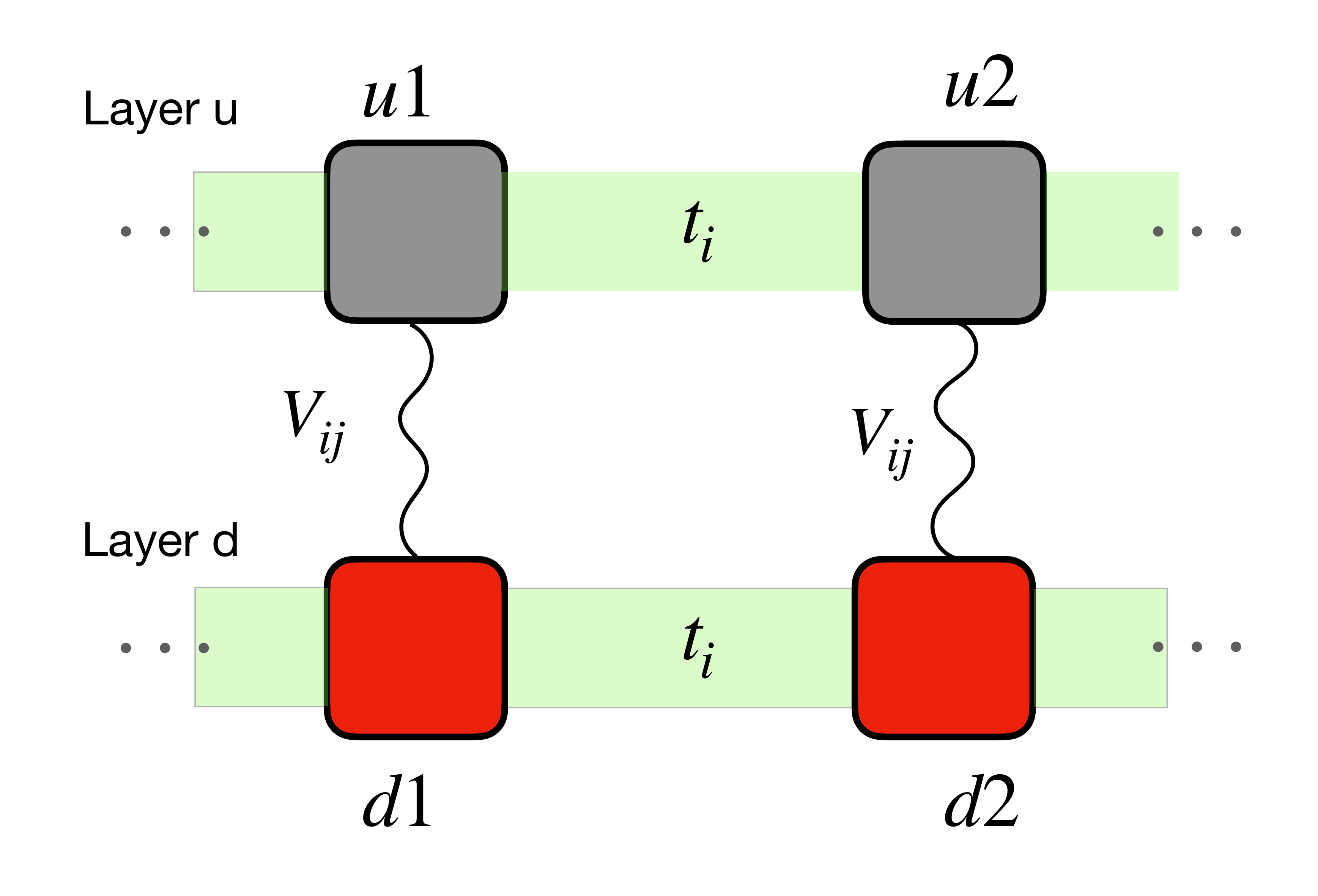}
  \caption{Schematic representation of the SYK double layer setup. The four depicted dots is the minimal set needed to evaluate the drag trans-conductance.}
  \label{fig:Device}
\end{figure} 

To reveal the main qualitative features of the  Coulomb drag in incoherent metals, we consider a theoretical model, which consists of two layers, dubbed by $u$ and $d$,  coupled by interactions. Each layer consists of an array of SYK dots, coupled by direct particle tunneling, see Fig.  \ref{fig:Device}. The Hamiltonian of the adopted model reads 
\begin{equation}
H=\sum_{\nu=u, d}\sum_{r} H_{\mathrm{SYK}}^{\nu r}+ \sum_{\nu=u,d} H_t^{\nu} + V_{\mathrm{int}}. 
\label{H}
\end{equation}
Here the first term describes the set of isolated SYK dots 
\begin{equation}
H_{\mathrm{SYK}}^{\nu r}=\sum_{ij, kl}^N J^{\nu r}_{ij, kl} c^+_{\nu r i}c^+_{\nu r j} c_{\nu r k} c_{\nu r l}, 
\label{Hsyk}
\end{equation} 
where $ J^{\nu r}_{ij, kl} $ are random couplings drawn from the Gaussian distribution with zero mean and the variances $\overline{|J^{\nu r}_{ij, kl}|^2}=\frac{2 J^2}{N^3}$. The interactions in different dots are statistically independent of each other. 
The second term in Eq. (\ref{H}) describes the inter-dot tunneling of electrons in each layer 
\begin{equation}
H_t^{\nu}=  \sum_{\langle r, r'\rangle}\sum_{i}^N t^{\nu}_i\left(c^+_{\nu r i}c_{\nu r' i}+h.c.\right),  
\label{Ht}
\end{equation} 
where $t^{\nu}_i$ denotes random tunneling amplitudes derived from the Gaussian distribution with zero mean and the variance $\overline{|t^{\nu}_i|^2}=t_0^2$, and the sum $\langle r, r'\rangle$ runs over the nearest neighbors. The tunneling couplings in different layers are statistically independent. We associate the site index $i,j,k,l$ within the SYK dot with a quantum number characterizing some quantum mechanical state (orbital), which is conserved by the tunneling. The last term in Eq. (\ref{H}) describes inter-layer interactions.  Being guided by the random interactions within the SYK dot, we adopt the random inter-dot interaction between the on-site charge densities   
 \begin{equation}
 V_{\mathrm{int}} =\sum_{i,j}^N V_{ i j} \sum_{r} c^+_{u r  i} c_{u r  i} c^+_{d r  j} c_{d r  j}. 
 \label{Vij}
 \end{equation}
 The random interaction constants $V_{ i j} $ have zero mean and are characterized by the variance $\left\langle V_{i j}^2\right\rangle=\frac{V^2}{N}$. 

Each isolated SYK dot provides a model of an incoherent metal that completely lacks electron or hole quasi-particles \cite{Sachdev1993, Kitaev2015}. However, a weak electron hopping within an array of SYK dots changes the low-energy spectrum, restoring coherent quasiparticles. This, in turn, induces a crossover between the high-temperature incoherent SYK metal and low-temperature FL metal at a temperature of $T_0\sim t_0^2/J$, which is determined by the electron escape rate from the SYK-grain \cite{Song2017,Feigelman2018,Chowdhury2018,ABK2019,Patel2019,Chowdhury2022}. The model adopted here exhibits the same crossover. As we will show, the crossover between the SYK and FL regimes results in a qualitative change in the Coulomb drag resistance.

\begin{figure}[t!]
  \centering
  \includegraphics[width=0.5\textwidth]{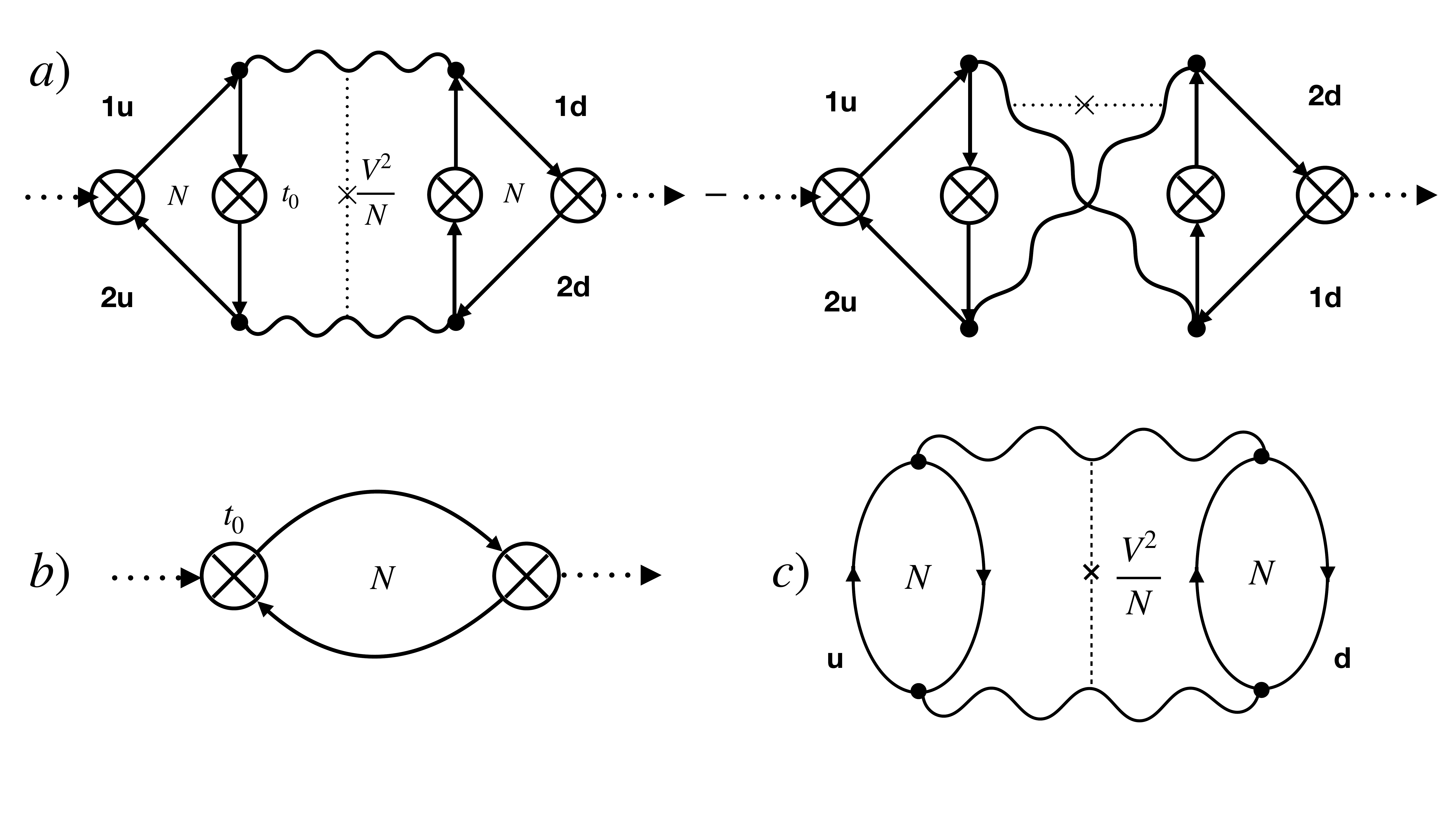}
  \vskip -.3cm
  \caption{a) Diagrams for the drag trans-conductance. Full lines represent interacting SYK Green functions, wavy lines -- inter-layer interactions, and crossed circles -- intra-layer tunneling; b) Diagram for the inter-dot conductance within a single layer. c) Diagram describing the heat current between the SYK dots in the up ($u$) and down ($d$) layers. Factors of $N$ are indicated explicitly.}
  \label{fig:Diagrams}
\end{figure}  

The trans-conductance  is determined by the drag conductance between the two 
 junctions, one from each layer, formed by the closest grains, which we assign with the numbers $r=1, r'=2$, as shown in Fig.  \ref{fig:Device}.    
The drag conductance is calculated according to the Kubo formula approach developed in Refs. \cite{Kamenev1995,Flensberg1995}. 
The basic diagram describing the drag response between the layers $u$ and $d$ is shown in Fig. \ref{fig:Diagrams}a).
Solid lines in Fig. \ref{fig:Diagrams} denote the one-particle Green's functions of the SYK model. Since Coulomb drag is possible only if the particle-hole symmetry is violated, we assume that the SYK grains in both layers are away from half-filling. The charge asymmetry is parametrized by the parameters, introduced for SYK model in Ref. \cite{Sachdev2015}, $\mathcal{E}_{u,d}\propto -d\mu_{u,d}/dT$, that are proportional to the temperature derivative of the corresponding chemical potentials. Detailed calculation outlined in the Supplemental Material \cite{Supplement} result in the following expression for the drag conductance in the strange metal (SM) regime
\begin{equation}
\sigma_{\mathrm{Drag}}^{\mathrm{SM}}\approx 58.6 N\frac{V^2}{ J^2} \frac{T_0^2}{T^2}\mathcal{E}_u \mathcal{E}_d.  
\label{sigma_approx}
\end{equation}
The drag conductance diminishes with temperature as $T^{-2}$. At the same time, the DC conductance within the layer behaves as $1/T$, \cite{Song2017}, 
$\sigma^{\mathrm{SM}} \approx 0.886 N\frac{e^2}{h} \frac{T_0}{T}$  (for details see Supplemental material \cite{Supplement}). 
As a result the drag trans-resistance between the two strange metals is {\em temperature-independent}: 
\begin{equation}
\rho_{\mathrm{Drag}}^{\mathrm{SM}}=\frac{\sigma_{\mathrm{Drag}}^{\mathrm{SM}}}{\left(\sigma^{\mathrm{SM}}\right)^2}\approx \frac{C^{\mathrm{SM}} }{N} \frac{h}{e^2} 
\frac{V^2}{J^2} \mathcal{E}_u \mathcal{E}_d, 
\label{Rd_SYK}
\end{equation}
where the numerical factor is estimated as $C^{\mathrm{SM}}\approx 74.7$, and the SM regime is realized for $T>T_0$. Furthermore, we find that in this regime drag resistance remains independent of the tunneling strength $t_0$. This universality leads us to conclude that the validity of Eq. (\ref{Rd_SYK}) extends beyond the specific microscopic model used in this paper. Besides the evident proportionality to the charge asymmetries in both layers, the only physical parameter governing the drag conductance is the ratio of the inter- and intra-layer interactions, $V/J$. Remarkably, this parameter can be independently determined through the measurement of the near-field heat transport \cite{Volokitin2007,Mahan2017,Jiang2017,Ben-Abdallah2021}, as we demonstrate below.

The temperature-independent drag resistance of the strange metal stands in stark contrast to the drag resistance in the Fermi liquid, which is proportional to $T^2$. 
In the Fermi liquid regime at $T<T_0<t_0$, tunneling is the most relevant term in the Hamiltonian. It smears the low-energy SYK singularity in the single-particle 
density of states, substituting it with a semi-circular energy band with a width of $4t_0$ (at larger energy, $2t_0 < \epsilon <J$, the SYK-like tails remain). Assuming that the two chemical potentials fall within this central band, $|\mu_{u,d}|< 2t_0$,   the calculations of the drag conductance that are outlined in the Supplemental Material \cite{Supplement} result in 
\begin{equation}
\sigma_{\mathrm{Drag}}^{\mathrm{FL}}\propto N\frac{V^2}{J^2} \frac{T^2}{T_0^2}\mathcal{E}_u \mathcal{E}_d.
\label{SigmaD_FL}
\end{equation}
 Meanwhile, the intra-layer conductance in the FL regime is independent of temperature, $\sigma^{\text{FL}}=\frac{e^2}{\pi h}N$. Therefore, the resulting drag resistance is given by
\begin{equation}
\rho^{\mathrm{FL}}_{\mathrm{Drag}}\approx \frac{C^{\mathrm{FL}}}{N} \frac{h}{e^2}\frac{V^2}{J^2}\frac{T^2}{T_0^2}\mathcal{E}_u\mathcal{E}_d,
\label{Rd_FL}
\end{equation}
where $C^{\mathrm{FL}}\approx 429.2$ (for a detailed derivation of these results, see the Supplemental Material \cite{Supplement}).

We conclude that the overall temperature dependence of the drag resistance rises as $\sim T^2$ at low temperatures in the Fermi liquid regime and saturates to a temperature independent value at high temperatures in the SM regime. The drag resistances given by Eqs. (\ref{Rd_SYK}), (\ref{Rd_FL}) become comparable in the range of temperature  $ T\sim T_0=t_0^2/J$ that marks the crossover between the Fermi-liquid and SM regimes. Since the numerical coefficient by the drag resistance in the Fermi liquid regime Eq. (\ref{Rd_FL}) is larger than the one in the SM regime Eq. (\ref{Rd_SYK}), the estimation of the drag resistance in the two regimes at $T=T_0$ gives $\rho^{\mathrm{FL}}_{\mathrm{Drag}}(T_0)>\rho^{\mathrm{SM}}_{\mathrm{Drag}}$, which suggests that the overall temperature dependence may exhibit a maximum at temperatures about $T_0$. 

\begin{figure}[t!]
  \centering
  \includegraphics[width=0.5\textwidth]{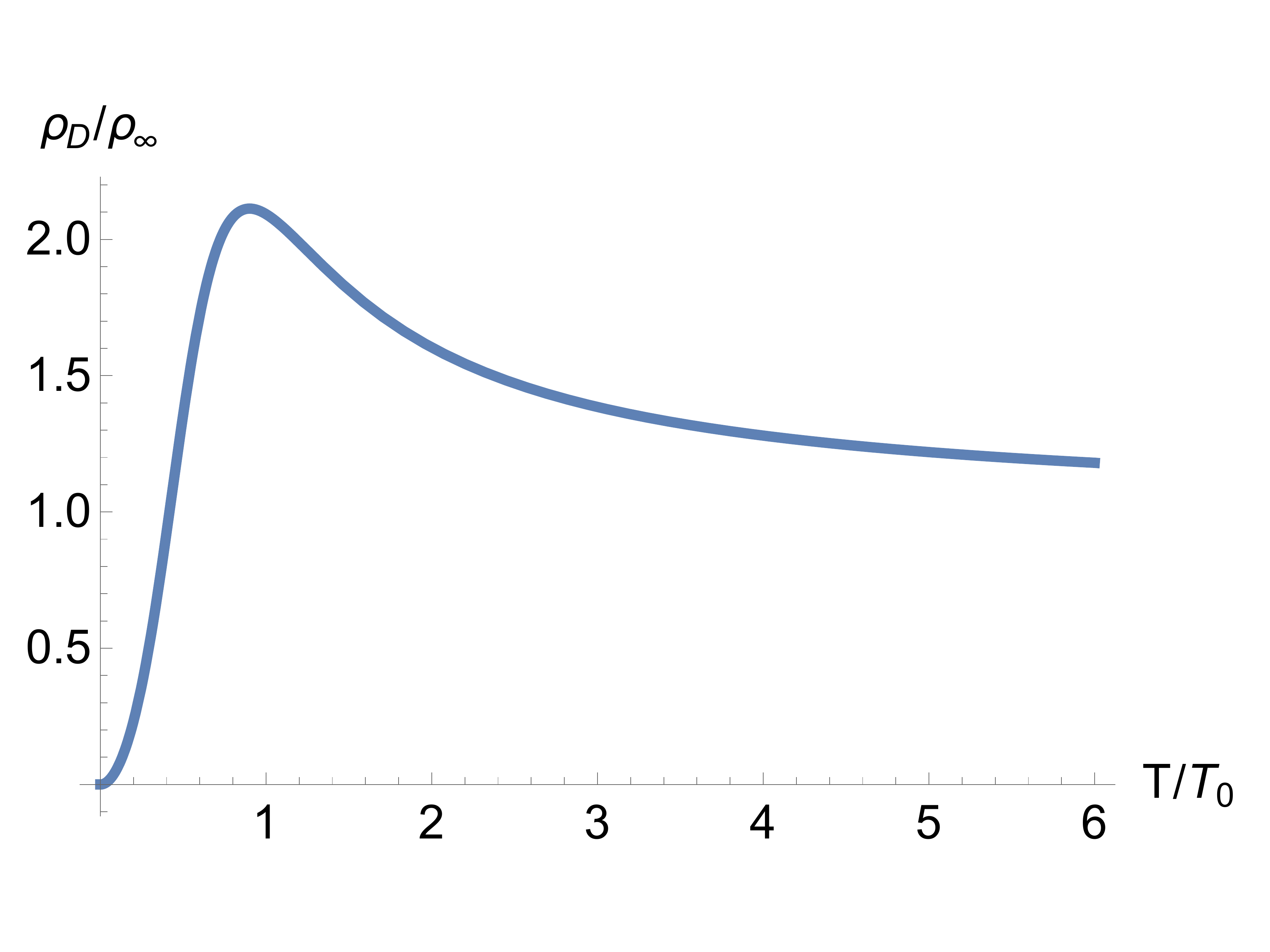}
  \vskip -.5cm
  \caption{Temperature dependence of the drag resistance (in units of the drag resistance at high temperature $\rho_{\infty}$): the $T^2$ increase of resistance in the low temperature FL regime changes to saturation in the high temperature SM regime.}
  \label{fig:Rd_T}
\end{figure}

One may derive a phenomenological expression for the overall temperature dependence of the drag resistance based on the following physical picture. The energy spectrum in the tunnel-coupled SYK dots can be roughly separated into two regions. The states within the energy window of the order of the tunneling escape rate $T_0=t_0^2/J$ form a quasi-Fermi liquid, contributing to the drag resistance according to Eq. (\ref{Rd_FL}). On the other hand, the energy states beyond the energy window of $T_0$ form the strange metal, leading to the drag resistance as given by Eq. (\ref{Rd_SYK}). Both parts of the spectrum constitute the two liquids, contributing in parallel to the overall resistance. Since the high-energy states' population necessitates their thermal activation, the two liquids' contributions should be weighted by their corresponding thermal activation probabilities, resulting in the following expression for the inverse resistance:
\begin{equation}
\frac{1}{\rho_{\mathrm{Drag}}}=\frac{1-e^{-T_0/T}}{\rho^{\mathrm{FL}}_{\mathrm{Drag}}}+\frac{e^{-T_0/T}}{\rho_{\mathrm{Drag}}^{\mathrm{SM}}}.
\label{empiric_Rdrag}
\end{equation}
Qualitative temperature dependence of the drag resistance is shown in Fig. \ref{fig:Rd_T}. 
It is important to note that the drag resistance calculation in the crossover regime necessitates exact form of the one-particle Green's functions of the tunnel-coupled SYK grains at the crossover temperature, which is currently unavailable to the best of our knowledge. Therefore, the question of whether the overall temperature dependence of the drag resistance exhibits a maximum remains open.

Consider now the near-field heat transfer conductance in the model described by Eqs. (\ref{H})--(\ref{Vij}). In the lowest order of interaction, the near-field heat transfer flux $J_h$ is given by the diagram shown in Fig. \ref{fig:Diagrams}c), leading to the following result for the heat conductance in the SM regime:
\begin{equation}
\varkappa^{\text{SM}}=\frac{J_h}{\Delta T} = 0.015 N \frac{V^2}{J^2} T, 
\label{Jh_result}
\end{equation}
where $T=(T_u+T_d)/2$, $\Delta T=T_u-T_d$, and we assume a small temperature difference $\Delta T\ll T$. Equation (\ref{Jh_result}) allows one to define the near-field heat conductance as $\varkappa^{\text{SM}}=J_h/\Delta T$, which is a linear function of temperature. The slope of the temperature dependence of the heat conductance is then directly related to the ratio $V^2/J^2$ characterizing the interaction strength in the SM regime. Therefore, one can relate the drag resistance and the heat conductance as follows 
\begin{equation}
\rho_{\mathrm{Drag}}^{\mathrm{SM}}=\frac{A^\mathrm{SM}}{N^2} \frac{h}{e^2} \mathcal{E}_u \mathcal{E}_d\frac{d\varkappa^{\text{SM}}}{dT}, 
\label{Rd_kappa}
\end{equation}
where the constant $A^\mathrm{SM}\approx 4980$. Eq. (\ref{Rd_kappa}) provides a universal relation between the results of two different experiments in the incoherent metal. 

Remarkably the same functional relation (\ref{Rd_kappa}) between the drag resistance and the heat conductance holds in the Fermi liquid regime with a somewhat different numerical coefficient $A^{\mathrm{FL}}\approx 180$. Indeed, the corresponding heat conductance is known to be \cite{Volokitin2007,Mahan2017,Jiang2017,Supplement} 
\begin{equation}
\varkappa^{\mathrm{FL}}=0.8 N \frac{V^2 T^3}{t_0^4}=0.8 N \frac{V^2}{J^2}\frac{T^3}{T_0^2}. 
\label{sigma_hFL}
\end{equation}
Along with Eq. (\ref{Rd_FL}) this leads to Eq.~(\ref{Rd_kappa}) with the aforementioned $A^{\mathrm{FL}}$. 

In summary, we studied the nonlocal electrical and thermal transport in the interactively coupled double-layers of two strange metals. Each layer is modeled by the Hamiltonian of tunnel-coupled SYK quantum dots. This model is known to capture the physics of strange metal phases in the proper regime of parameters. If the temperature is smaller than the characteristic scale set by inter-grain tunneling and intra-grain interaction, we recover the FL regime with the quadratic temperature dependence of drag resistivity [Eq. \eqref{Rd_FL}]. In the temperature range above that scale, we find trans-resistance approaching the limiting value, Eq. \eqref{Rd_SYK}, from above. The latter fact reflects the interplay of Planckian intra-layer dissipation and interaction-mediated inter-layer dragging. Results obtained for our microscopic model differ from the recent study of the drag between two strange metal layers using the Einstein-Maxwell-dilaton model from holography, which claims $\rho_{\text{Drag}}\propto T^4$  \cite{Stoof}. Finally, we calculated near-field inter-layer thermal conductance. The established relationship, Eq. \eqref{Rd_kappa}, between drag resistance and the near-field heat conductance that is free of parameters of the considered model suggests universality of this result.  

We thank A. Patel for the communication regarding Ref. \cite{Patel2019}. This work at UW-Madison was financially supported by the National Science Foundation Grant No. DMR-2203411 (A.L.).  A.K. was supported by the NSF Grant No. DMR-2037654. A.C. thanks the Fine Theoretical Physics Institute at the University of Minnesota for hospitality and support.

\bibliography{SYK_drag.bib}

\newpage
\onecolumngrid

\section{Coulomb drag and heat transfer in strange metals: Supplemental Material}

In this Supplemental Material we provide details of the derivation of important formulas in the main text of the paper. In order to maintain a coherent and comprehensive presentation, we repeat here the Hamiltonian and the main diagrams (see Fig. \ref{fig:Diagram_Drag}) for the calculation of the drag conductance and the near field heat transfer. 
The Hamiltonian of the two SYK-array layers is given by   
\begin{equation}
H=\sum_{\nu=u, d}\sum_{r} H_{\mathrm{SYK}}^{\nu, r}+ \sum_{\nu=u,d} H_t^{\nu} + V_{\mathrm{int}}. 
\label{Hs}
\end{equation}
The Hamiltonian of an isolated SYK grain reads 
\begin{equation}
H_{\mathrm{SYK}}^{\nu, r}=\sum_{ij, kl}^N J^{\nu r}_{ij, kl} c^+_{\nu r i}c^+_{\nu r j} c_{\nu r k} c_{\nu r l}, 
\label{Hsyks}
\end{equation} 
where $J^{\nu r}_{ij, kl} $ are random couplings drown from Gaussian distribution with zero mean and the variances $\overline{|J^{\nu r}_{ij, kl}|^2}=\frac{2 J^2}{N^3}$. The interactions in different grains are statistically independent of each other. 
The inter-grain tunneling of an  electron in a single layer is governed by the Hamiltonian 
\begin{equation}
H_t^{\nu}=  \sum_{\langle r,r'\rangle } \sum_{i} t_i\left(c^+_{\nu r i}c_{\nu r' i}+h.c.\right) 
\label{Hts}
\end{equation} 
where $t_i$ denote random tunneling amplitudes derived from the Gaussian distribution with zero mean and the variance $\overline{|t_i|^2}=t_0^2$, and $\langle r, r'\rangle$ denotes a pair of the nearest neighbor grains.
The random inter-grain interaction between the on-site charge densities is given by 
 \begin{equation}
 V_{\mathrm{int}} =\sum_{i,j} V_{ i j} \sum_{r} c^+_{u r  i} c_{u r  i} c^+_{d r  j} c_{d r  j},  
 \label{Vijs}
 \end{equation}
 where the random interaction constants $V_{ i j} $ have zero mean and are characterized by the variance $\left\langle V_{i j} V_{kl}\right\rangle=\frac{V^2}{N}$.

\subsection{One particle Green functions in the strange metal (high-temperature) regime}

Since Coulomb drag is possible only if the particle-hole symmetry is violated, we assume that the SYK grains in both layers are away from half-filling. The charge asymmetry is parametrized by the parameters $\mathcal{E}_u$ and $\mathcal{E}_d$ correspondingly. Below we write down the single particle Green functions in the imaginary time $\tau$ and in the Matsubara frequency representations \cite{Sachdev2015,Chowdhury2022}
\begin{eqnarray}
&& G(\tau)= -\frac{\pi^{1/4}\sqrt{T} e^{-2\pi \mathcal{E} T \tau}}{\sqrt{2J \sin(\pi T \tau)}} , \, \, \, (\tau>0), \label{Gtau>}\\ 
&& G(\tau)= \frac{\pi^{1/4}\sqrt{T} e^{-2\pi\mathcal{E}-2\pi \mathcal{E} T \tau}}{\sqrt{2J \sin(-\pi T \tau)}} , \, \, \, (\tau<0), \label{Gtau<}\\ 
&& G(i \omega_n)=-i\frac{C(\mathcal{E})}{\sqrt{2\sqrt{\pi } J T}}\frac{\Gamma\left(\frac{3}{4}+n+i\mathcal{E}(\theta)\right)}{\Gamma\left(\frac{5}{4}+n+i\mathcal{E}(\theta)\right)}, \label{G_iomega} 
\end{eqnarray}
where $\omega_n=2\pi T(n+1/2)$, and the constants $C$, $\mathcal{E}$ determine the charge asymmetry. They are related to each other as follows 
\begin{equation}
C(\mathcal{E})=\frac{(1+i)\left(1+i e^{-2\pi \mathcal{E}}\right) }{2i}. 
\end{equation}
Analytical continuation to real frequencies results in the following retarded and advanced Green functions 
\begin{equation}
G^R(\omega)=\left(G^{A}(\omega)\right)^* =\frac{-i e^{-i\theta}}{\sqrt{2 J T}\left(\pi\cos(2\theta)\right)^{1/4}} \frac{\Gamma\left[1/4-i\left(\frac{\omega}{2\pi T}-\mathcal{E}\right)\right]}{\Gamma\left[3/4-i\left(\frac{\omega}{2\pi T}-\mathcal{E}\right)\right]}. \label{G_Romega}
\end{equation}
Here the phase factor $\theta$, $-\pi/4 < \theta <\pi/4$ relates to the charge asymmetry parameters $\mathcal{E}$ and $C$ as follows 
\begin{eqnarray}
&& 
C(\theta) =\frac{\pi^{1/4}}{\sqrt{J}\left[\cos(2\theta)\right]^{1/4}}, \label{C} \\ 
&&
\mathcal{E}(\theta)=\frac{1}{2\pi}\ln\left(\frac{1+\tan\theta}{1-\tan\theta}\right) \, \, \, <=> \, \, \, 
\tanh(2\pi \mathcal{E}) = \tan(\theta) . \label{E}
\end{eqnarray} 
 $\theta=0$ corresponds to the charge symmetry point (the half-filling). Note also, that in the SYK model,  the charge asymmetry parameter relates to the chemical potential as 
 \begin{equation}
 \mu_0-\mu=2\pi T\mathcal{E},
 \label{muE}
 \end{equation} 
 where $\mu_0$ denotes the chemical potential at zero temperature \cite{Sachdev2015,davison2017thermoelectric,Chowdhury2022}. 
 In what follows we use the expression for the Green function of the dimensionless frequency $x=\frac{\omega}{2\pi T}$ 
 \begin{equation}
 G^{R/A}(\omega) =K_{\mathrm{syk}} g^{R/A}_{\mathrm{syk}}(x), 
 \label{G_SYK}
 \end{equation}
 where
\begin{eqnarray}
&& 
K_{\mathrm{syk}}=\frac{1}{\sqrt{2\sqrt{\pi} JT}}, \label{K_syk}\\ 
&& 
g^{R}_{\mathrm{syk}}(x)=\left(g^{A}_{\mathrm{syk}}(x)\right)^* =\frac{-i e^{-i\theta}}{\left(\cos(2\theta)\right)^{1/4}} \frac{\Gamma\left[1/4-i\left(x -\mathcal{E}\right)\right]}{\Gamma\left[3/4-i\left(x -\mathcal{E}\right)\right]}. \label{g_syk}
\end{eqnarray} 

\subsection{One particle Green functions in the Fermi liquid (low-temperature) regime}
At temperatures less then single-electron tunneling rate between the two SYK-grains, $T<T_0=t_0^2/J$, the array of grains enters the Fermi liquid regime.  In that regime, the transport properties of the array are determined by the energies less than $T_0$. In turn, the single particle spectrum of each grain at those energies is determined by random inter-grain tunneling amplitudes, which leads to the semi-circular energy band with the width  $4 t_0$ given by the variance of the random tunneling. Close to the center of the band, $|\omega+\mu|\ll 2 t_0$,  the Green functions can be written as 
\begin{equation}
G^R(\omega)\approx \frac{2}{\omega+\mu+\frac{i}{2\tau_0}},  
\label{GRtau}
\end{equation}
where the life time $\tau_0$ is determined by the bandwidth   
\begin{equation}
\frac{1}{2 \tau_0}= 2t_0.
\label{tau0}
\end{equation} 
The advanced Green function is given by the complex conjugated expression.  Similarly to the SYK regime discussed above, we further  use the Green function of dimensionless frequency $x=\frac{\omega}{2\pi T}$
 \begin{equation}
 G^{R/A}(\omega) =K_{\mathrm{fl}} g^{R/A}_{\mathrm{fl}}(x).
  \label{G_FL}
 \end{equation}
Here 
\begin{eqnarray}
&& 
K_{\mathrm{fl}}=\frac{1}{\pi T}, \label{K_fl}\\ 
&& 
g^{R}_{\mathrm{fl}}(x)=\left(g^{A}_{\mathrm{fl}}(x)\right)^* =\frac{1}{x -\mathcal{E} +\frac{i}{w}},  \label{g_fl}
\end{eqnarray} 
where the dimensionless parameter $w$ corresponding to the decay time is defined as $w=4\pi T \tau_0=\pi T/t_0$, and $\mathcal{E}=-\mu/(2\pi T)$.

\begin{figure}[htb]
  \centering
  \includegraphics[width=\textwidth]{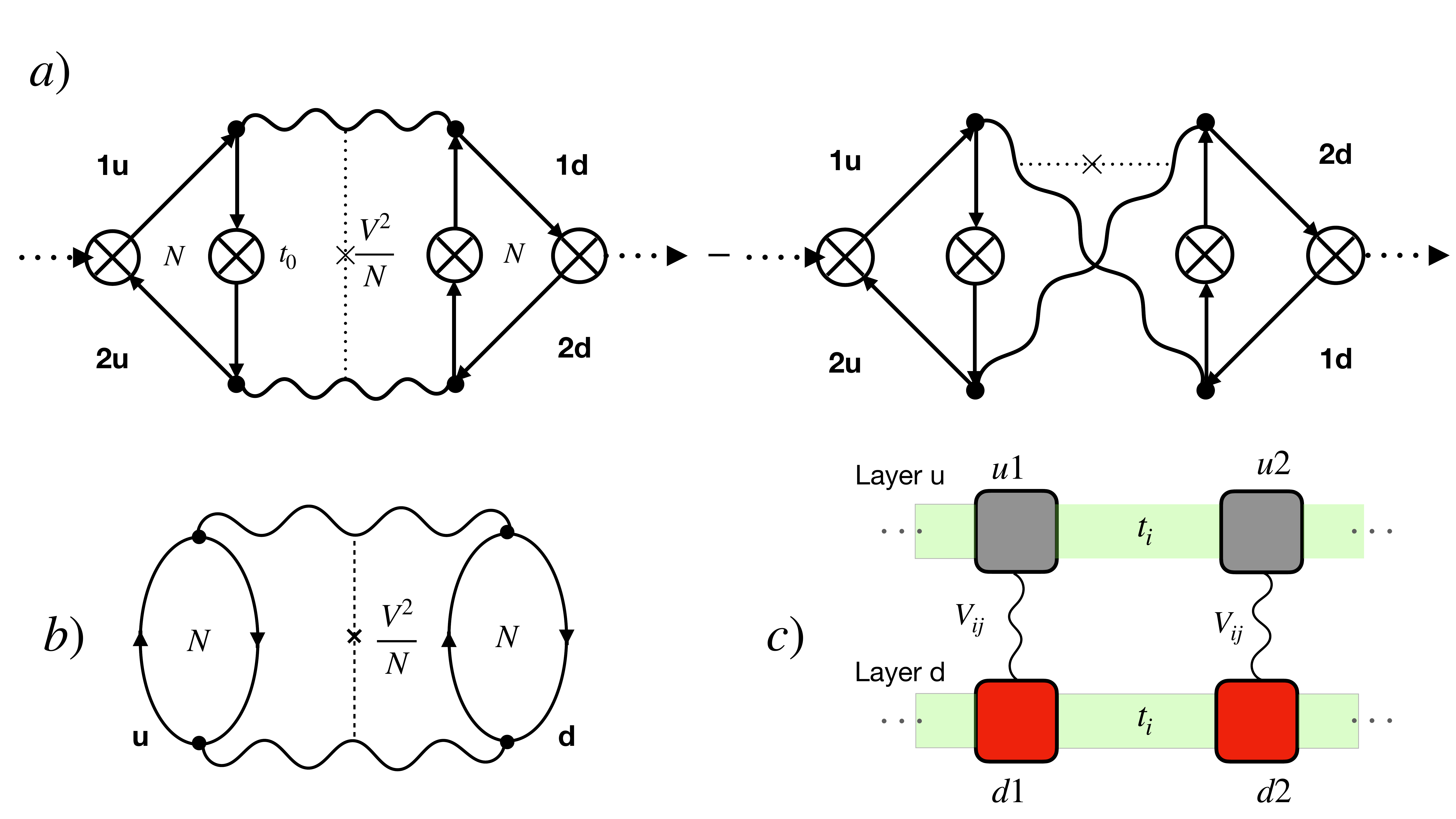}
  \caption{a) Diagrams for the drag response; b) Diagram for the near field heat transfer response; c) Scheme of the two layer system.}
  \label{fig:Diagram_Drag}
\end{figure}  
 
\section{Calculation of drag conductance}

Here we calculate the drag conductance according to the Kubo formula approach developed in Ref. \cite{Kamenev1995}. 
\begin{equation}
\sigma_{\mathrm{Drag}}=\frac{V^2}{16\pi T N}\int_{-\infty}^{\infty} \frac{d\omega}{\sinh^2\left(\frac{\omega}{2T}\right)} \Gamma_u^{+-}(\omega, \omega) \Gamma_d^{+-}(\omega, \omega). 
\label{Kubo_KamenevOreg}
\end{equation}
The basic diagrams describing the drag-response between the layers $u$ and $d$ are shown in Fig. \ref{fig:Diagram_Drag}a). Solid lines in Fig. \ref{fig:Diagram_Drag} denote the one-particle Green functions of the SYK model. The factors 
 $\Gamma_{u,d}^{+-}(\omega, \omega) $ denote the triangular parts of the diagram in Fig. \ref{fig:Diagram_Drag}a), each  corresponding to the mathematical expression 
\begin{eqnarray}
\nonumber && 
\Gamma^{+-}(\omega,\omega)=\frac{N t_0^2}{4\pi i}\int d\epsilon \left[\tanh\left(\frac{\epsilon+\omega}{2T}\right) - \tanh\left(\frac{\epsilon}{2T}\right)\right] \left[(G^A(\epsilon))^2 - (G^R(\epsilon))^2\right] G^R(\epsilon+\omega)G^A(\epsilon+\omega) \\ 
&& 
-\left\{\omega \rightarrow -\omega\right\} .
\label{Gamma+-}
\end{eqnarray}
The one particle Green functions in Eq. (\ref{Gamma+-}) should be taken with the charge asymmetry parameters 
$\mathcal{E}_{u/d}$ for the up and down layer respectively. 
Introducing dimensionless frequencies  $\xi=\frac{\omega}{2\pi T}$ 	and $x=\frac{\epsilon}{2\pi T}$  we can cast the expression for 
$\Gamma^{+-}(\omega,\omega)$ to the form 
\begin{equation}
\Gamma^{+-}(\omega,\omega)=-\frac{i}{2} N t_0^2 T K^4 \left[\gamma(\xi)-\gamma(-\xi)\right], 
\label{Gamma_gamma}
\end{equation}
where 
\begin{equation}
\gamma(\xi)=\int d x \left[\tanh\left(\pi (x+\xi)\right) - \tanh\left(\pi x \right)\right] \left[(g^A(x))^2 - (g^R(x))^2\right] g^R(x+\xi)g^A(x+\xi), 
\label{gamma_xi}
\end{equation}
where the constant $K$ and the dimensionless Green functions $g(x)$ are determined by Eqs. (\ref{K_syk}), (\ref{gSYK}) and Eqs. (\ref{K_fl}), (\ref{g_fl}) in the strange metal (SM) and Fermi liquid (FL) regimes respectively. 

Then,  using dimensionless frequencies, Eq. (\ref{Kubo_KamenevOreg}) can be rewritten in the form 
\begin{equation}
\sigma_{\mathrm{Drag}}=-\frac{N V^2 t_0^4}{32}T^2 K^8 \int_{-\infty}^{\infty} \frac{d\xi}{\sinh^2\left(\pi \xi \right)} \left[\gamma_u(\xi)-\gamma_u(-\xi)\right]\left[\gamma_d(\xi)-\gamma_d(-\xi)\right].
\label{sigmaD_gamma}
\end{equation}

\subsection{Small-$\mathcal{E}$ expansion of the drag conductance}

The lowest order term in the small - $\mathcal{E}$ expansion of the drug conductance is obtained from Eq. (\ref{sigmaD_gamma}) as 
\begin{equation}
\sigma_{\mathrm{Drag}}\approx -\mathcal{E}_u\mathcal{E}_d \frac{N V^2 t_0^4}{32}T^2 K^8 
\int_{-\infty}^{\infty} \frac{d\xi}{\sinh^2(\pi \xi)}\left[\gamma'(\xi)-\gamma'(-\xi)\right]^2, 
\label{sigma_smallE}
\end{equation}
where 
\begin{equation}
\gamma'(\xi)=\frac{\partial\gamma(\xi)}{\partial\mathcal{E}}\bigg|_{\mathcal{E}=0}. 
\label{gamma'}
\end{equation}

\subsection{Drag conductance in the SM regime}
Substituting explicit expression Eq. (\ref{g_syk}) in Eq. (\ref{gamma_xi}), we obtain 
\begin{eqnarray}
\nonumber 
\gamma(\xi)=\int dx \frac{\left[\tanh(\pi(x+\xi))-\tanh(\pi x)\right]}{\cos(2\theta)} \left[e^{2 i \theta}\left(\frac{\mathrm{\Gamma}(1/4+i(x-\mathcal{E}))}{\mathrm{\Gamma}(3/4+i(x-\mathcal{E}))}\right)^2-e^{-2 i \theta} \left(\frac{\mathrm{\Gamma}(1/4-i(x-\mathcal{E}))}{\mathrm{\Gamma}(3/4-i(x-\mathcal{E}))}\right)^2\right] \\ 
\nonumber 
\times  \frac{\mathrm{\Gamma}(1/4-i(x+\xi-\mathcal{E}))}{\mathrm{\Gamma}(3/4-i(x+\xi-\mathcal{E}))} \frac{\mathrm{\Gamma}(1/4+i(x+\xi-\mathcal{E}))}{\mathrm{\Gamma}(3/4+i(x+\xi-\mathcal{E}))}. \\ 
\label{gamma}
\end{eqnarray}
For the explicit calculation of derivative over $\mathcal{E}$ it is convenient to shift the integration variable $x\rightarrow x+\mathcal{E}$ in Eq. (\ref{gamma}).  Then, in the SM regime we obtain 
\begin{eqnarray}
\nonumber && 
\gamma'(\xi)=\int_{-\infty}^{\infty} dx \left[\frac{\pi}{\cosh^2[\pi(x+\xi)]}-\frac{\pi}{\cosh^2(\pi x)}\right]\left[(g^R_{\mathrm{syk}}(x))^2-(g^A_{\mathrm{syk}}(x))^2\right]
g^R_{\mathrm{syk}}(x+\xi)g^A_{\mathrm{syk}}(x+\xi) + \\ 
&& 
4\pi i \int_{-\infty}^{\infty} dx \left[\tanh[\pi(x+\xi)]-\tanh(\pi x)\right] \left[(g^R_{\mathrm{syk}}(x))^2+(g^A_{\mathrm{syk}}(x))^2\right]g^R_{\mathrm{syk}}(x+\xi)g^A_{\mathrm{syk}}(x+\xi), 
\label{gamma'syk}
\end{eqnarray}
where the functions $g^{R/A}_{\mathrm{syk}}(x)$ are taken for $\mathcal{E}=0$, 
\begin{equation}
g^R_{\mathrm{syk}}(x)=\left(g^A_{\mathrm{syk}}(x)\right)^*=\frac{\mathrm{\Gamma}(1/4-ix)}{\mathrm{\Gamma}(3/4-i x)}
\label{gSYK}
\end{equation}
Furthermore, evaluating the derivative over $\mathcal{E}$, and using Eq. (\ref{sigma_smallE}) together with Eq. (\ref{K_syk}), we obtain the expression for the drag conductance in the final form 
\begin{equation}
\sigma_{\mathrm{Drag}}^{\mathrm{SM}}\approx \mathcal{C} \frac{N t_0^4 V^2}{ J^4 T^2} \mathcal{E}_u \mathcal{E}_d= 
\mathcal{C} N \frac{V^2}{ J^2}\left(\frac{T_0}{T}\right)^2 \mathcal{E}_u \mathcal{E}_d , 
\label{sigma_approxs}
\end{equation}
where the constant $\mathcal{C}$ is evaluated numerically as $\mathcal{C}\approx 58.6$, and we introduced the crossover temperature $T_0=t_0^2/J$ in the second equation.

\subsection{One particle conductance in the SM regime}

We calculate one particle conductance as the tunneling conductance between the two SYK grains according to the formula 
\begin{equation}
\sigma_1^{\mathrm{SM}}= \frac{e^2}{\hbar} N t_0^2 \int \frac{d\omega}{2\pi} \frac{\nu_{\mathrm{SYK}}^2(\omega)}{4T\cosh^2\left(\frac{\omega}{2T}\right)}, 
\end{equation} 
where $\nu_{\mathrm{SYK}}$ denotes the one particle density of states in the SYK grain at the Fermi energy, which can be obtained from the imaginary part of the one particle Green function Eq. (\ref{G_Romega}). Since we calculate the drag resistance in the lowest order of the charge asymmetry parameter $\mathcal{E}$, the calculation of the one particle conductance can be performed for the charge symmetric point $\mathcal{E}=0$. Then we obtain 
\begin{equation}
\nu_{\mathrm{SYK}}=-\frac{1}{\pi} \mathrm{Im} G^R(\omega, \mathcal{E}=0)=\frac{\sqrt{2}}{\pi^{1/4}\sqrt{JT}}\mathrm{Re}
\left[\frac{\Gamma\left(1/4-i\frac{\omega}{2\pi T}\right)}{\Gamma\left(3/4-i\frac{\omega}{2\pi T} \right)} \right],
\end{equation}
\begin{equation}
\sigma_1^{\mathrm{SM}} =\frac{e^2}{h}  \frac{N t_0^2}{2\sqrt{\pi} JT} \int_{-\infty}^{\infty}\frac{dx}{\cosh^2(\pi x)} \left(\mathrm{Re}\frac{\Gamma(1/4-ix)}{\Gamma(3/4-ix)} \right)^2\approx 0.886 \frac{e^2}{h} \frac{N t_0^2}{JT}=0.886 N \frac{e^2}{h} \frac{T_0}{T} ,  
\end{equation}
which is in accord with results of Ref. \cite{Song2017}

\subsection{Drag resistance in the SM regime}
The drag resistance is obtained as 
\begin{equation}
\rho_{\mathrm{Drag}}^{\mathrm{SM}}=\frac{\sigma_{\mathrm{Drag}}^{\mathrm{SM}}}{\left(\sigma_1^{\mathrm{SM}}\right)^2}\approx C_{\mathrm{SM}}  \frac{h}{e^2} 
\frac{V^2}{N J^2} \mathcal{E}_u \mathcal{E}_d, 
\label{Rd_SYKs}
\end{equation}
where the numerical factor $C_{\mathrm{SM}}$ is estimated as $C_{\mathrm{SM}}\approx 74.7$. 
Therefore, the SM drag resistance is independent of temperature.

\subsection{Drag conductance in the Fermi liquid regime}
It is the chemical potential $\mu$ rather than the dimensionless parameter $\mathcal{E}$ that determines the filling fraction in the Fermi liquid regime.  For small charge asymmetry (close to the half-filling), the chemical potential is much smaller than the random energy bandwidth, $\mu\ll 4t_0$. In contrast to the linear temperature dependence of the chemical potential in the SM regime, in the FL regime the chemical potential $\mu$ at a constant filling is only weakly dependent on temperature. Yet for the sake of technical convenience we use Eqs. (\ref{K_fl}), (\ref{g_fl}), formulated in terms of dimensionless quantities for the evaluation of the drag conductance in the Fermi liquid regime. We note that despite $\mathcal{E}=-\mu/(2\pi T)$ can become large for temperatures close to zero, it still remains much smaller than $1/w=t_0/(\pi T)$. It follows that the product $ \mathcal{E} w =-\mu/(2t_0)$ plays the role of the small parameter for the expansion close to the charge-symmetry point (half-filling). 

Substituting Eq. (\ref{g_fl}) in Eq. (\ref{gamma_xi}), and shifting the integration variable $x-\mathcal{E}\rightarrow x$ we obtain 
\begin{eqnarray}
\nonumber && 
\gamma(\xi)-\gamma(-\xi)=\int_{-\infty}^{\infty} \frac{4ix/w}{\left(x^2+\frac{1}{w^2}\right)^2} \left\{ \left[\tanh[\pi(x+\mathcal{E}+\xi)]-\tanh[\pi(x+\mathcal{E})]\right] \frac{1}{(x+\xi)^2+\frac{1}{w^2}} - \right. \\ 
&& 
\left. 
\left[\tanh[\pi(x+\mathcal{E}-\xi)]-\tanh[\pi(x+\mathcal{E})]\right] \frac{1}{(x-\xi)^2+\frac{1}{w^2}}\right\}. 
\label{gamma-gamma}
\end{eqnarray}
Furthermore, changing the integration variable $x\rightarrow -x$ in the second line, and rescaling $x\rightarrow \frac{x}{w}$, we cast Eq. (\ref{gamma-gamma}) into the form suitable for the expansion in  small  $w\mathcal{E}$ 
\begin{eqnarray}
\nonumber && 
\gamma(\xi)-\gamma(-\xi)=\\ 
\nonumber && 
\int_{-\infty}^{\infty} dx \frac{4 i w^3 x}{(x^2+1)^2}\frac{1}{(x+w\xi)^2 +1} \left\{ 
\tanh\left[\frac{\pi}{w}(x+w\xi+w\mathcal{E})\right]- \tanh\left[\frac{\pi}{w}(x+w\xi-w\mathcal{E})\right]- \right. \\  
\nonumber && 
\left. 
\left(\tanh\left[\frac{\pi}{w}(x+w\mathcal{E})\right]-\tanh\left[\frac{\pi}{w}(x-w\mathcal{E})\right]\right) \right\} \approx \\ 
\nonumber && 
8 i \pi w^3 \mathcal{E}\int_{-\infty}^{\infty} \frac{x dx}{(x^2+1)^2[(x+ w\xi)^2+1]} \left[\frac{1}{\cosh^2\left[\frac{\pi}{w}(x+ w\xi)\right]}  - \frac{1}{\cosh^2\left(\frac{\pi}{w}x\right)}\right] \approx \\ 
 && 
8 i \pi w^3 \mathcal{E} \left\{\int_{-w/\pi}^{w/\pi} \frac{(-w\xi) dx}{(w^2\xi^2+1)^2(x^2+1)} - \int_{-w/\pi}^{w/\pi} 
\frac{x dx}{(x^2+1)^2}\frac{1}{w^2\xi^2+1}\right\} =-i\frac{16 w^5 \mathcal{E} \xi}{(1+w^2\xi^2)^2}. 
\label{gamma_wE}
\end{eqnarray}

Substituting Eq. (\ref{gamma_wE}) in Eq. (\ref{sigmaD_gamma}), we obtain the integral over $\xi$ in the leading order in $w$ in the form 
\begin{equation}
\int_{-\infty}^{\infty} \frac{d\xi}{\sinh^2(\pi \xi)}\left[\gamma'(\xi)-\gamma'(-\xi)\right]^2=  -2^{8} w^{10} \int_{-\infty}^{\infty} \frac{d\xi}{\sinh^2(\pi \xi)}\frac{\xi^2}{(1+w^2\xi^2)^2}\approx -\frac{2^{8}}{3\pi} w^{10}. 
\label{xi-integral_FL}
\end{equation}

Finally, restoring all factors in Eq. (\ref{sigma_smallE}),  using  $w=\pi T \tau_0=\pi T/t_0$ in Eq. (\ref{xi-integral_FL}), and taking into account the relation between $\mathcal{E}$ and the chemical potential $\mu$,  we obtain 
\begin{equation}
\sigma_{\mathrm{Drag}}^{\mathrm{FL}}\approx  \frac{2}{3} \frac{N V^2 T^2}{t_0^6}\mu_u\mu_d.
\label{sigmaDrag_FL}
\end{equation}

\subsection{One particle conductance in the Fermi liquid regime}
The one particle conductance is calculated as the tunneling conductance between the nearest neighbor grains, using the simplified expression for the Green functions Eq. (\ref{GRtau}), which results in \cite{Song2017}
\begin{equation}
\sigma_1 =N \frac{e^2}{\pi h}. 
\end{equation}
Therefore, the one particle conductance is independent of tunneling strength and temperature (at low temperatures). The independence of the one-particle conductance of the tunneling strength is explained by the fact that the bandwidth (and thus the one-particle density of states) is determined by the variance of the randomized  inter-grain tunneling, which leads to the exact compensation between the one particle density of states and the inter-grain tunneling amplitude. 

\subsection{Drag resistance in the Fermi liquid regime}

The drag resistance is obtained as 
\begin{equation}
\rho^{\mathrm{FL}}_{\mathrm{Drag}}=\frac{\sigma_{\mathrm{Drag}}^{\mathrm{FL}}}{\sigma_1^2}\approx \frac{2 \pi^2}{3}  \frac{h}{e^2}\frac{V^2 T^2}{N t_0^6}\mu_u\mu_d. 
\label{Rd_FLs}
\end{equation}
Therefore, the drag resistance in the Fermi liquid regime is proportional to the temperature squared.

\subsection{Drag conductance in the crossover regime}
Here we compare the values of the drag conductance in the SM and in the FL regime at the crossover temperature $T_0=t_0^2/J$. We show that the estimations for the two regimes differ at the crossover temperature only by the numerical factor. We assume that the charge asymmetry remains constant for all temperatures. 

In the SM regime, the charge asymmetry is determined by the asymmetry factor $\mathcal{E}$. At small charge asymmetry (close to the half-filled system), the deviation from the half filling is given by the expression \cite{Sachdev2015,Chowdhury2022} 
\begin{equation}
\mathcal{Q}-1/2\approx -\mathcal{E}(1+\pi/2).
\label{Q_SYK}
\end{equation}

In the Fermi liquid regime, the charge asymmetry is determined by the chemical potential. Close to the half filling at low temperature, the deviation from the half filling is given by 
\begin{equation}
\mathcal{Q}-1/2=\int d\omega \nu_1(\omega)\left(\frac{1}{e^{(\omega-\mu)/T}+1}-\frac{1}{2}\right)\approx \nu_1(0) \mu, 
\end{equation}
where $\nu_1(\omega)$ denotes the single-particle density of states. The one particle density of states (DOS) $\nu_1(0)$ at the middle of the semicircular  band is given by 
\begin{equation}
\nu_1(0)=4\tau_0/\pi=1/(\pi t_0),
\end{equation}
which leads to the relation 
 \begin{equation}
\mathcal{Q}-1/2\approx \frac{\mu}{\pi t_0}. 
\label{Q_FL} 
\end{equation}

Equating the expressions  for the filling in the strange metal and in the Fermi liquid regime given by Eqs. (\ref{Q_SYK}) and (\ref{Q_FL}), we obtain the relation between the chemical potential in the Fermi liquid and the asymmetry parameter in the SYK, which reads 
\begin{equation}
\mu^{\mathrm{FL}}=- \pi t_0 (1+\pi/2) \mathcal{E}^{\mathrm{SM}}. 
\label{muFL_E}
\end{equation}
Now we use Eq. (\ref{muFL_E}) to compare the estimations for the drug conductance in the SM and in the FL regimes at the temperature $T_0=t_0^2/J$. In the FL regime we can represent Eq. (\ref{sigmaDrag_FL}) in the form 
\begin{equation}
\sigma_{\mathrm{Drag}}\approx  \frac{2}{3} N\pi^2 (1+\pi/2)^2 \frac{V^2 T^2}{t_0^4}\mathcal{E}_u\mathcal{E}_d =
  \frac{2}{3} N \pi^2 (1+\pi/2)^2 \frac{V^2}{J^2}\left(\frac{T}{T_0}\right)^2\mathcal{E}_u\mathcal{E}_d .
\label{sigmaDrag_FL_TT0}
\end{equation}
At $T=T_0$ we obtain 
\begin{equation}
\sigma^{\mathrm{FL}}_{\mathrm{Drag}}(T_0)\approx 43.5 N  \frac{V^2}{J^2}\mathcal{E}_u \mathcal{E}_d.  
\label{sigmaDrag_FLT0}
\end{equation}
In the SM regime we obtain from Eq. (\ref{sigma_approx}) 
\begin{equation}
\sigma^{\mathrm{SM}}_{\mathrm{Drag}}(T_0)\approx 58.6 N  \frac{V^2}{J^2}\mathcal{E}_u \mathcal{E}_d.  
\label{sigmaDrag_SYKT0}
\end{equation}
One can see that the estimation for the drag conductance at the crossover temperature $T_0$ differ by the numerical prefactor only. 

Furthermore, expressing the chemical potentials through the spectral asymmetry parameters in Eq. (\ref{Rd_FL}), and introducing the crossover temperature $T_0=t_0^2/J$ we obtain the drag resistance in the Fermi liquid regime in the form  
\begin{equation}
\rho^{\mathrm{FL}}_{\mathrm{Drag}}\approx \frac{C_{\mathrm{FL}}}{N} \frac{h}{e^2}\frac{V^2}{J^2}\frac{T^2}{T_0^2}\mathcal{E}_u\mathcal{E}_d,
\label{Rd_FLEs}
\end{equation}
where $C_{\mathrm{FL}}\approx 429.2$

\section{Calculation of near field heat transfer flux}
In this section we calculate the near field heat transport between the up and down layers. We assume that the heat transport takes place between the nearest grains in the different layers, that is between the grains with the same number $(1d\leftrightarrow 1u)$,  $(2d\leftrightarrow 2u)$. 
The general expression for the near field heat transfer flux, illustrated by the diagram Fig. \ref{fig:Diagram_Drag}b), is given by \cite{KamenevBook2} 
\begin{equation}
J_h= \frac{V^2}{N} \int\frac{d\omega}{2\pi} (\mathrm{Im} \Pi^R_u (\omega)) (\mathrm{Im}\Pi^R_d (\omega)) \omega \left[n_B\left(\omega/T_u\right)-n_B\left(\omega/T_d\right)\right]. 
\label{HeatCurrent}
\end{equation}
Here $n_B(\omega/T)$ denotes the Bose distribution at temperature $T$,  and $\mathrm{Im} \Pi^R_{u/d} (\omega)$ denotes the imaginary part of the polarization operator. Since no charge asymmetry is required for a finite heat transfer, we perform calculations at the charge symmetric point $\mathcal{E}_{u}=\mathcal{E}_{d}=0$. 

We calculate the polarization operator using the Keldysh formalism, where it is defined as \cite{KamenevBook2} 
\begin{equation}
\Pi^R (\omega)=\frac{i}{2}N\int\frac{d\epsilon}{2\pi} \left\{G^R(\epsilon+\omega)G^K(\epsilon)+G^K(\epsilon+\omega)G^A(\epsilon)\right\}. 
\label{defPi}
\end{equation}
Here $G^K(\epsilon)=\tanh\left(\frac{\epsilon}{2T}\right)\left[G^R(\epsilon)-G^R(\epsilon)\right]$ denotes the Keldysh Green function at temperature $T$.

Introducing dimensionless frequencies $x=\frac{\epsilon}{2\pi T}$, $\xi=\frac{\omega}{2\pi T}$, and using the definitions of the Green functions Eqs. (\ref{G_SYK}), (\ref{G_FL}),  we cast Eq. (\ref{defPi}) in the form 
\begin{equation}
\Pi^R (\omega)= \frac{i}{2} N K^2 T\kappa(\xi), 
\label{Pi_kappa}
\end{equation} 
where 
\begin{equation}
\kappa(\xi)=\int_{-\infty}^{\infty} dx \left[g^R(x+\xi)+g^A(x-\xi)\right]\left[g^R(x)-g^A(x)\right]\tanh(\pi x). 
\label{kappa}
\end{equation}
For the  further evaluation we assume small temperature difference between the layers, $T_u=T-(\Delta T)/2$, $T_d=T+(\Delta T)/2$. 
Then, substituting Eqs. (\ref{Pi_kappa}), (\ref{kappa}) in Eq. (\ref{HeatCurrent}) and expanding the difference of Bose distribution functions in $\Delta T$, we obtain 
\begin{equation}
J_h=\frac{\pi^2}{8} N V^2 K^4 T^3(\Delta T) \int_{-\infty}^{\infty} d\xi \frac{\xi^2 \left[\mathrm{Re}\, \kappa(\xi)\right]^2}{\sinh^2(\pi \xi)}. 
\label{Jh_final}
\end{equation}

\subsection{Strange metal regime}
Substituting $K=K_{\mathrm{syk}}$ as given by Eq. (\ref{K_syk}) and $g^{R/A}= g_{\mathrm{syk}}^{R/A}$ as given by Eq. (\ref{g_syk}) in Eq. (\ref{Jh_final}), we obtain
\begin{equation}
J_h^{\mathrm{SM}}=\frac{\pi N V^2}{16 J^2}T(\Delta T) \int d\xi \frac{\xi^2 \left[\mathrm{Re}\, \kappa(\xi)\right]^2}{\sinh^2(\pi \xi)}. 
\label{Jh_finalSYK}
\end{equation}
Further numerical evaluation of the integral in Eq. (\ref{Jh_final}) results in the expression for the heat transfer flux 
\begin{equation}
J_h^{\mathrm{SM}}\approx 0.015 \frac{N V^2}{J^2}T(\Delta T), 
\end{equation}
from which it follows for the heat conductance 
\begin{equation}
\varkappa_h^{\mathrm{SM}}\approx 0.015 \frac{N V^2}{J^2}T. 
\label{sigma_hs}
\end{equation}

\subsection{Fermi liquid regime}
In the Fermi liquid regime we use Eqs. (\ref{K_fl}) and (\ref{g_fl}). Substituting them into Eq. (\ref{Jh_final}), we obtain 
\begin{equation}
I_h^{\mathrm{FL}}=\frac{N V^2}{4\pi^2} \frac{\Delta T}{T} \int_{-\infty}^{\infty} d\xi \frac{\xi^2 \left[\mathrm{Re}\, \kappa(\xi)\right]^2}{\sinh^2(\pi \xi)}
\label{Jh_FL}
\end{equation}
Since there is no scale invariance of the Green functions in the FL regime, the last integration does not reduce to a number. Rather it is a function of the dimensionless density of states in the center of the random energy band $w=\pi T/t_0$. Explicitly we obtain for $\kappa(\xi)$ from Eq. (\ref{kappa}) 
\begin{equation}
\mathrm{Re} \kappa^{\mathrm{FL}}(\xi)=\frac{1}{2} w^2\xi \int_{-\infty}^{\infty} dy \frac{y \tanh(2\pi y/w)}{y^2+1/4} \frac{1}{[y^2-\frac{1}{4}(w\xi)^2+\frac{1}{4}]^2+\frac{1}{4}(w\xi)^2}.
\label{kappa_FL}
\end{equation}
For the further evaluation we note, that $T\ll t_0$ in the FL regime hence $w\ll 1$, and that the actual values of $\xi$ are restricted by $|\xi|\lesssim 1$ by the following integration over $\xi$ in Eq. (\ref{Jh_final}). Those conditions justify the evaluation of $\kappa^{\mathrm{FL}}$ in the leading order in $w$ and $\xi$, which contributes to the neglecting the terms $w\xi$ in the denominator of Eq. (\ref{kappa_FL}) and replacing $\tanh(2\pi y/w)\approx \mathrm{sign}(y)$. Under those approximations we obtain 
\begin{equation}
\mathrm{Re} \kappa^{\mathrm{FL}}(\xi) \approx  w^2\xi \int_0^{\infty} \frac{y dy}{(y+1/4)^3}=4 w^2\xi. 
\label{kappa_FLfinal}
\end{equation}
Substituting Eq. (\ref{kappa_FLfinal}) in Eq. (\ref{Jh_FL}) and performing integration over $\xi$, we obtain the heat transfer flux in the FL regime in the form 
\begin{equation}
J_h^{\mathrm{FL}}\approx 0.8 N \frac{V^2 T^3}{t_0^4} \Delta T, 
\label{Jh_finalFL}
\end{equation}
from which we deduce the heat conductance 
\begin{equation}
\varkappa^{\mathrm{FL}}=0.8 N \frac{V^2 T^3}{t_0^4}=0.8 N \frac{V^2}{J^2}\frac{T^3}{T_0^2}, 
\label{sigma_hFLs}
\end{equation}
where in the last equation we introduced the crossover temperature $T_0=t_0^2/J$ between the SM and the FL regimes. 

Comparison of Eqs. (\ref{sigma_hs}) and (\ref{sigma_hFLs}) leads to conclusion  
\begin{equation}
\varkappa^{\mathrm{FL}}\propto \varkappa^{\mathrm{SM}} \left(\frac{T}{T_0}\right)^2. 
\label{SYK-FL_heat}
\end{equation}

\end{document}